\documentclass[preprint2]{aastex61}
\usepackage{float,graphicx,amsmath,multirow}
\usepackage{color}

\begin{document}

\title{Detection of Interstellar HC$_5$O in TMC-1 with the Green Bank Telescope}
\author{Brett A. McGuire}
\altaffiliation{B.A.M. is a Jansky Fellow of the National Radio Astronomy Observatory}
\affiliation{National Radio Astronomy Observatory, Charlottesville, VA 22903, USA}
\affiliation{Harvard-Smithsonian Center for Astrophysics, Cambridge, MA 02138, USA}
\author{Andrew M. Burkhardt}
\affiliation{Department of Astronomy, University of Virginia, Charlottesville, VA 22904, USA}
\author{Christopher N. Shingledecker}
\affiliation{Department of Chemistry, University of Virginia, Charlottesville, VA 22904, USA}
\author{Sergei V. Kalenskii}
\affiliation{Astro Space Center, Lebedev Physical Institute, Russian Academy of Sciences, Moscow, Russia}
\author{Eric Herbst}
\affiliation{Department of Chemistry, University of Virginia, Charlottesville, VA 22903, USA}
\affiliation{Department of Astronomy, University of Virginia, Charlottesville, VA 22903, USA}
\author{Anthony J. Remijan}
\affiliation{National Radio Astronomy Observatory, Charlottesville, VA 22903, USA}
\author{Michael C. McCarthy}
\affiliation{Harvard-Smithsonian Center for Astrophysics, Cambridge, MA 02138, USA}
\affiliation{School of Engineering and Applied Sciences, Harvard University, Cambridge, MA 02138, USA}
\correspondingauthor{Brett A. McGuire}
\email{bmcguire@nrao.edu}

\begin{abstract}
\noindent We report the detection of the carbon-chain radical HC$_5$O for the first time in the interstellar medium toward the dark cloud TMC-1 using the 100 m Green Bank Telescope.  We observe four hyperfine components of this radical in the $J = 17/2 \rightarrow 15/2$ rotational transition that originates from the $^2\Pi_{1/2}$ fine structure level of its ground state, and calculate an abundance of $n/n_{H_2}$ = $1.7\times 10^{-10}$, assuming an excitation temperature of $T_{ex} = 7$~K.  No indication of HC$_3$O, HC$_4$O, HC$_6$O, is found in these or archival observations of the source, while we report tentative evidence for HC$_7$O.  We compare calculated upper limits, and the abundance of HC$_5$O to predictions based on (1) the abundance trend of the analogous HC$_n$N family in TMC-1 and (2) a gas-grain chemical model.  We find that the gas-grain chemical model well reproduces the observed abundance of HC$_5$O, as well as the upper limits of HC$_3$O, HC$_6$O, and HC$_7$O, but HC$_4$O is over produced. The prospects for astronomical detection of both shorter and longer HC$_n$O chains are discussed.
\end{abstract}
\keywords{Astrochemistry, ISM: molecules, ISM: individual objects (TMC-1)}

\section{Introduction}
\label{intro}

Observations of complex chemistry occurring outside the typical hot core environments are critical for understanding the underlying reaction mechanisms and chemical evolutionary processes at work in the larger interstellar medium (ISM). One of the prototypical sources for such investigations is the dark cloud TMC-1, which contains a rich chemical inventory distinct from star-forming regions.  Indeed, while most hot core sources display a wealth of saturated organic molecules\footnote{Those species which have few double and triple carbon-carbon bonds and instead use these electrons to bind hydrogen atoms.} such as methanol (CH$_3$OH), ethanol (CH$_3$CH$_2$OH), dimethyl ether (CH$_3$OCH$_3$), and ethyl cyanide (CH$_3$CH$_2$CN) \citep{Neill:2014cb}, the inventory in TMC-1 is heavily weighted toward \emph{un}saturated species such as HC$_n$N ($n$~=~1$-$9) \citep{Loomis:2015jh}, C$_n$H ($n$~=~3--6), C$_2$S, C$_3$S, and C$_3$O \citep{Kaifu:2004tk}.   Since hydrogenation reactions are much more efficient on grains (e.g. CO $\rightarrow$ CH$_3$OH), saturated species will tend to be predominantly found on grain surfaces at the low temperatures within cold cores ($\le$20 K) \citep{Charnley:1995od,Garrod:2013id}. Therefore, gas-phase reactions (such as carbon insertion processes) that proceed rapidly at low temperature and density will tend to dominate the chemistry in these regions, resulting in the efficient production of unsaturated hydrocarbons and other carbon-chain molecules \citep{Herbst:2008to}.

Because cold cores such as TMC-1 are at an early stage of stellar evolution, their relatively simple physical history and well-defined conditions allow one to model and test chemical pathways, often in great detail. For species that are observed in both cold and hot cores, it is then possible to study the effects of the vastly differing physical conditions (temperature, density, radiation field, etc.) on the chemical evolution. Despite the relative dearth of complex organic molecules, another important characteristic of cold cores is the ease with which new molecular species can be unambiguously identified, especially at cm-wavelengths, due their cold excitation conditions, narrow linewidths, and uncrowded ($\sim$1~line per 200~km~s$^{-1}$) spectrum. For these reasons, it is not surprising that TMC-1 is one of the most well-studied astrochemical sources outside of hot cores and the carbon-star IRC+10216;  several dozen new molecular detections have been reported there over the past several decades (see \citet{Kaifu:2004tk} for an extensive review).

We have recently conducted high-sensitivity observations of TMC-1 in search of a number of new molecules.  Here, we report the first of several new detections from this study: the carbon-chain HC$_5$O radical via observation of four hyperfine components in the $J = 17/2 \rightarrow 15/2$ rotational transition.  The observations are presented in \S\ref{observations}, a review of the laboratory spectroscopy of HC$_5$O in \S\ref{spectroscopy}, the results and analysis in \S\ref{results}, and a discussion of the astrochemical implications and future study is given in \S\ref{discussion}.

\section{Observations}
\label{observations}

The observations were conducted over eight observing sessions from 2017 February to 2017 June using the 100 m Robert C. Byrd Green Bank Telescope in Green Bank, WV. The observations of TMC-1 were centered on $\alpha$(J2000)~=~04$^h$41$^m$42.5$^s$, $\delta$(J2000)~=~25$^{\circ}$41$^{\prime}$27.0$^{\prime\prime}$.  Pointing observations were conducted every hour; the pointing accuracy is estimated to be within 2$^{\prime\prime}$.  The K-band Focal Plane Array was used with the VEGAS spectrometer backend configured to provide 187.5 MHz total bandwidth in each of ten spectrometer banks at a 1.4 kHz (0.02 km s$^{-1}$) spectral resolution.  These extremely high-resolution observations were necessary to resolve the $\sim$0.3~km~s$^{-1}$ FWHM spectral features typical of TMC-1 \citep{Kaifu:2004tk}. The total spectral coverage was 1875 MHz in ten discontinuous 187.5 MHz windows within the range of 18 to 24 GHz.

Observations were conducted in position-switching mode, using a 1$^{\circ}$ offset throw, with 120 s of integration at each position and between $\sim$7.5 and 15~hours of total on-source integration depending on the frequency window.  The resulting spectra were placed on the atmosphere-corrected $T_A$* scale \citep{Ulich:1976yt}.  Data reduction was performed using the GBTIDL software package.  The spectra were averaged using a weighting scheme which corrects for the measured value of $T_{sys}$ during each 240 s ON-OFF cycle.  The spectra were smoothed to a resolution of 5.7 kHz (0.08~km~s$^{-1}$), sufficient to provide $\geq$3 points across each 0.3~km~s$^{-1}$ FWHM.  A polynomial fit was used to correct for baseline fluctuations.  The final RMS noise in the HC$_5$O window examined here was 2.4~mK.

\section{Spectroscopy}
\label{spectroscopy}

The pure rotational spectrum of HC$_5$O was precisely measured between 6 and 26 GHz by \citet{Mohamed:2005ku}.  This radical was produced in an electrical discharge of HC$_4$H + CO, and its spectrum measured using a Balle-Flygare cavity Fourier-Transform microwave spectrometer \citep{Balle:1981ex}.  Rest frequencies were determined to better than 1 ppm (2~kHz;~0.03~km~s$^{-1}$ at 20 GHz).  The ground electronic state of HC$_5$O is $^2\Pi_{r}$ with the $^2\Pi_{1/2}$ fine structure level lying lowest in energy, many tens of K below the $^2\Pi_{3/2}$ level.  At high spectral resolution, its rotational spectrum displays well-resolved $\Lambda$-doubling but more closely spaced hydrogen hyperfine splitting.   At the low rotational temperature characteristic of TMC-1, only  transitions from the lower $^2\Pi_{1/2}$ ladder are significantly populated; those that fall within the frequency coverage of the observations are given in Table~\ref{transitions}. Of the ten spectrometer windows used in these observations, only one, covering the range of 21766 to 21953 MHz contained HC$_5$O transitions.

\begin{table*}
\centering
\footnotesize
\caption{Measured and observed frequencies of HC$_5$O transitions covered in this work as well as pertinent line parameters.}
\begin{tabular}{c c c c c c c c c c}
\hline\hline
$J^{\prime} \rightarrow J^{\prime\prime}$	& $F^{\prime} \rightarrow F^{\prime\prime}$	&	$e/f$		&	Frequency (meas.)$^{a}$	&	Frequency (obs.)$^{b}$	&	Diff.		&	$\Delta T_A ^{c}$	&	$\Delta V^{d}$		&	$S_{ij}\mu^2$	&	$E_{u}$	\\
								& 									&			&	(MHz)				&	(MHz)				&	(kHz)	&	(mK)			&	(km~s$^{-1}$)	&	(Debye$^2$)	&	(K)		\\
\hline
17/2 $\rightarrow$ 15/2				& 9 $\rightarrow$ 8						&	$e$		&	21941.846			&	21941.848			&	-1		&	13.0			&	0.30			&	41.7			&	5.017	\\
								& 8 $\rightarrow$ 7						&	$e$		&	21941.977			&	21941.980			&	-2		&	13.6			&	0.27			&	37.0			&	5.018	\\
								& 9 $\rightarrow$ 8						&	$f$		&	21945.232			&	21945.231			&	2		&	15.1			&	0.26			&	41.7			&	5.019	\\
								& 8 $\rightarrow$ 7						&	$f$		&	21945.370			&	21945.370			&	1		&	11.0			&	0.33			&	37.0			&	5.018	\\
\hline
\multicolumn{10}{l}{$^{a}$\citet{Mohamed:2005ku}; 1$\sigma$ experimental uncertainty is $\sim$2 kHz.}\\
\multicolumn{10}{l}{$^{b}$Gaussian fit to line at $v_{lsr}$~=~5.64~km~s$^{-1}$.  1$\sigma$ uncertainty from Gaussian fit is $\sim$0.5 kHz.  Given the SNR}\\
\multicolumn{10}{l}{of the detected lines ($\sim$5.4) and the linewidth, we estimate the uncertainty in the observed line centers to be $\sim$3.7 kHz.}\\
\multicolumn{10}{l}{$^{c}$We estimate a conservative uncertainty of 30\% in the overall flux calibration.}\\
\multicolumn{10}{l}{$^{d}$Uncertainty in Gaussian fit is $\sim$0.01~km~s$^{-1}$.}\\
\end{tabular}
\label{transitions}
\end{table*}

\section{Results and Analysis}
\label{results}

\begin{figure*}
\centering
\includegraphics[width=\textwidth]{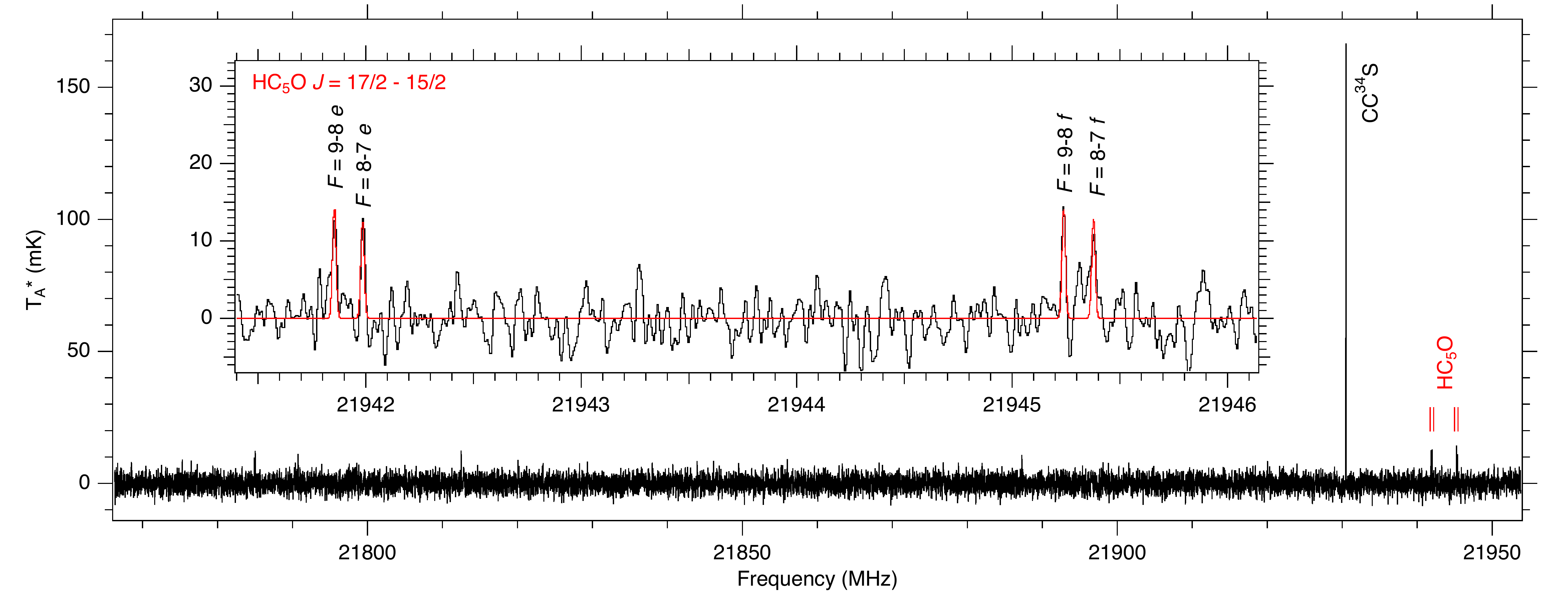}
\caption{Spectrum (\textbf{black}) toward TMC-1 in the frequency range containing the HC$_5$O transitions.  The inset provides an expanded view of the HC$_5$O features.  A simulation of the radical at $T_{ex}$~=~7~K from the laboratory work of \citet{Mohamed:2005ku}, at a linewidth of 0.26~km~s$^{-1}$ and a v$_{lsr}$~=~5.64~km~s$^{-1}$ is overlaid in red.  The quantum numbers for each observed transition are labeled.}
\label{spectra}
\end{figure*}

We observe emission from four hyperfine components of HC$_5$O in the $J=17/2 \rightarrow 15/2$ rotational transition at $v_{lsr} = 5.64$~km~s$^{-1}$, typical of molecules in this source \citep{Kaifu:2004tk}; the parameters of the observed lines are given in Table~\ref{transitions} and the spectra toward TMC-1 are shown in Figure~\ref{spectra}.  Although these lines originate from a single $J$ rotational level, taken together, the $\Lambda$-doubling and hydrogen hyperfine splitting provide a unique spectroscopic signature.  Given the very low line density of the spectra and the coincidence of the line centers to less than the experimental uncertainties, a mis-identification of these lines is extremely unlikely from these data alone.

Molecular emission in TMC-1 has been shown to be well-modeled by a single excitation temperature \citep{Remijan:2006tw}, typically $T_{ex} =$~7--10~K.  Because of the negligible spread ($\sim$0.1\%) in upper-state energies probed by the observed transitions, we assume a $T_{ex} = 7$~K for our abundance determination.  The column density of HC$_5$O was determined using the formalism of \citet{Hollis:2004uh}, given in Eq.~\ref{column}, 
\begin{equation}
N_T = \frac{Qe^{E_u/T_{ex}}}{\frac{8\pi^3}{3k}\nu S\mu^2}\times \frac{\frac{1}{2}\sqrt{\frac{\pi}{\ln(2)}}\frac{\Delta T_A\Delta V}{\eta_B}}{1 - \frac{e^{h\nu/kT_{ex}} -1}{e^{h\nu/kT_{bg}} -1}}
\label{column}
\end{equation}
where $N_T$ is the column density (cm$^{-2}$), $E_u$ is the upper state energy (K), $\Delta T_A \Delta V$  is integrated line intensity (K~cm~s$^{-1}$), $T_{ex}$ is the excitation temperature (K), $T_{bg}$ is the background continuum temperature (2.7 K), $\nu$ the transition frequency (Hz), $S$ is the intrinsic line strength, $\mu^2$ is the transition dipole moment (Debye)\footnote{These units must be properly converted to Joules$\cdot$cm$^3$ to give $N_T$ in cm$^{-2}$.}, and $\eta_B$ is the beam efficiency ($\sim$0.7 for the GBT at 20 GHz). The rotational partition function, $Q$, is calculated explicitly by direct summation of states ($Q$[7~K]~=~491). We assume that the source fills the beam (see \citet{Loomis:2015jh} for a detailed discussion).

\subsection{HC$_{\textup{\emph{5}}}$O}

For HC$_5$O, we calculate a column density of $1.7$~$\times$~$10^{12}$~cm$^{-2}$ at $T_{ex} = 7$~K. Assuming the HC$_5$O is co-spatial with previous estimates of H$_2$ in the region [$N$(H$_2$)~=~10$^{22}$~cm$^{-2}$; \citet{Gratier:2016fj}], this corresponds to an abundance of $n/n_{H_2}$~=~$1.7$~$\times$~$10^{-10}$.  \vspace{2em}

\subsection{HC$_{\textup{\emph{3}}}$O, HC$_{\textup{\emph{4}}}$O, HC$_{\textup{\emph{6}}}$O, HC$_{\textup{\emph{7}}}$O}
\label{ulims}

\begin{figure}
\centering
\includegraphics[width=0.5\textwidth]{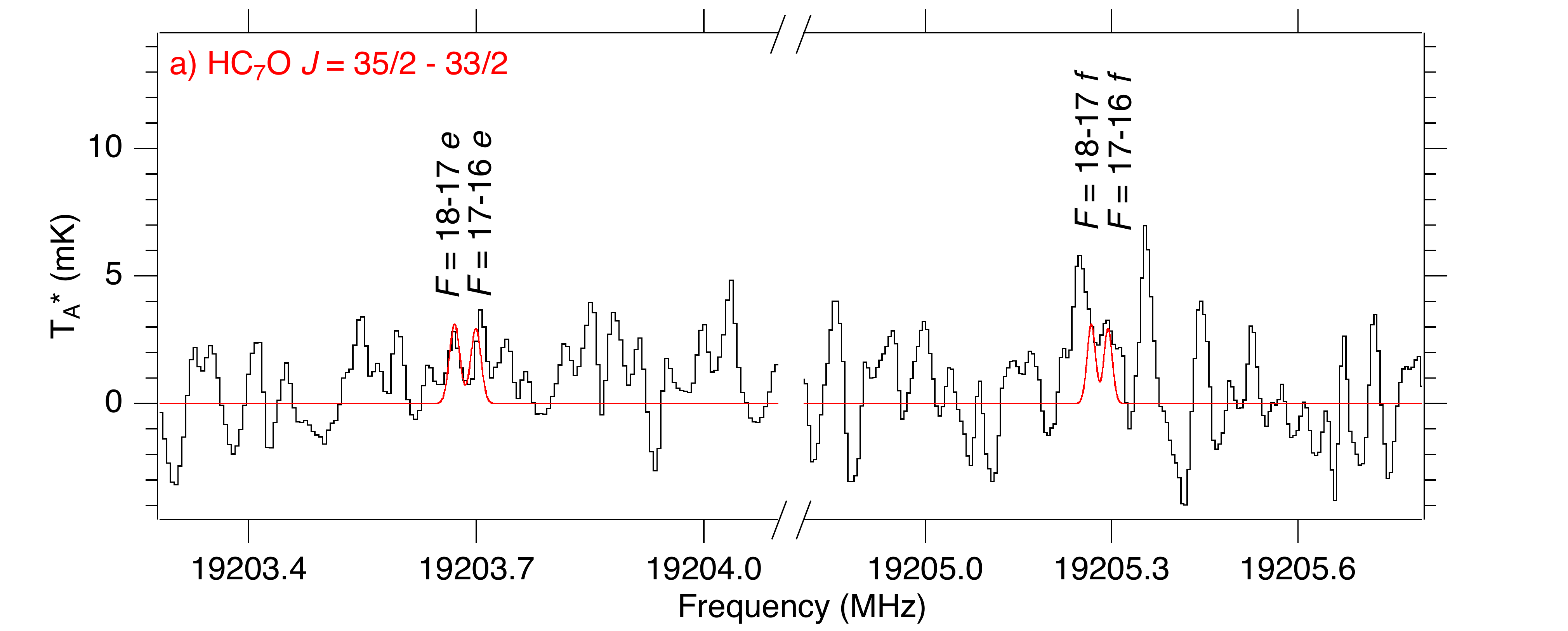}
\includegraphics[width=0.5\textwidth]{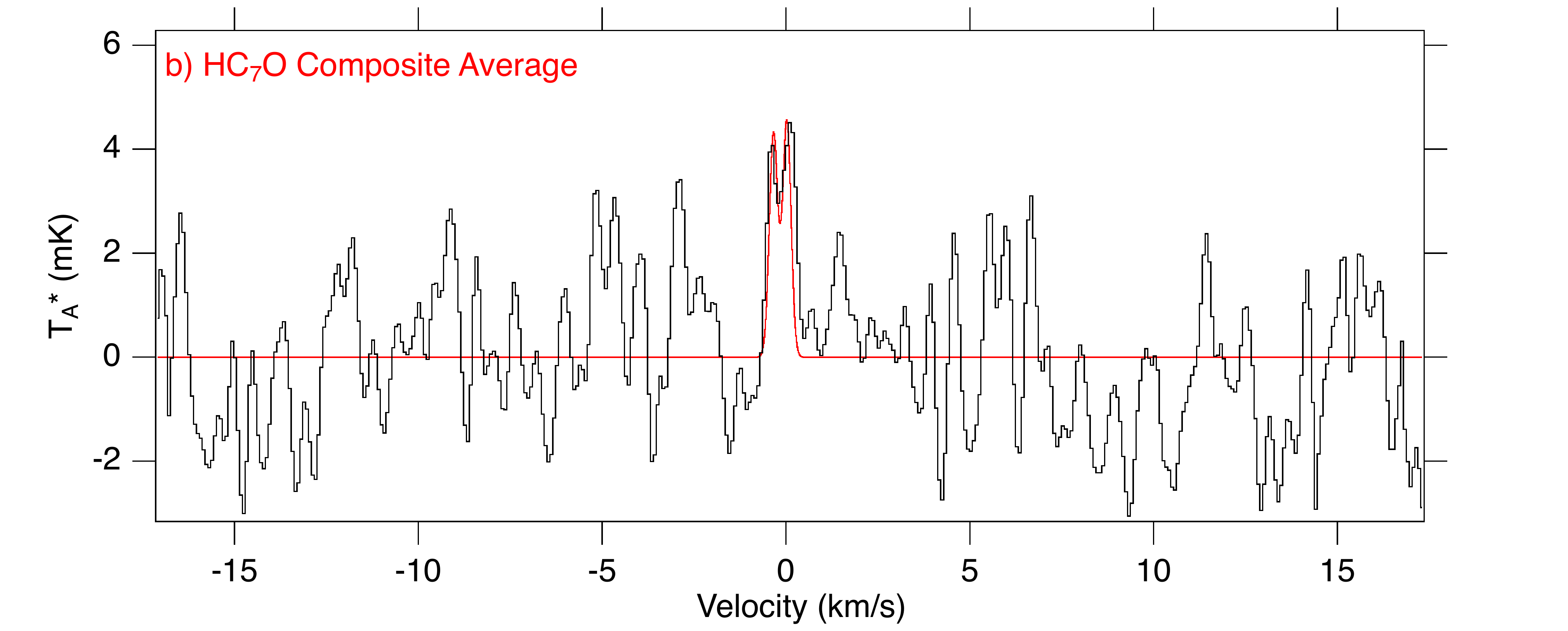}
\caption{A) observational spectra coincident with the $J = 35/2 - 33/2$ rotational transition of HC$_7$O in black, with the HC$_7$O transitions overlaid in red.  This is the lowest RMS noise window containing an HC$_7$O line, and there is no detection of HC$_7$O.  B) Composite average of the $J = 35/2 - 33/2$ and $J = 37/2 - 35/2$ transitions of HC$_7$O; because the hydrogen hyperfine splitting is not fully resolved, the average was performed using the central frequency, preserving the partial splitting.  The SNR of the (tentative) composite line is $\sim$3.}
\label{hc7o}
\end{figure}

We have also searched our observations, and those of \citet{Kaifu:2004tk}, for other members of the HC$_n$O ($n$~=~3--7) family of oxygen-terminated hydrocarbon free radicals.   At present, our observations only cover transitions of HC$_7$O , and no detection is seen in individual transitions (Fig.~\ref{hc7o})a.  If the four $\Lambda$-doubling components of the two $J$ transitions covered in our observations are stacked in velocity space using a composite average approach \citep{Kalenskii:2010hj}, some tentative indication of HC$_7$O is seen, suggesting the individual transitions remain at or just below our detection limit (Fig.~\ref{hc7o}b).

We do not find any definitive evidence in our survey, or that of \citet{Kaifu:2004tk}, for signal arising from any of the other HC$_n$O radicals.   Assuming the same $T_{ex} = 7$~K as for HC$_5$O, we have calculated upper limits to the column densities using the strongest transition in the available frequency coverage (Table~\ref{columns}).  \citet{Mohamed:2005ku} lists the calculated dipole moments for the  HC$_n$O radicals discussed here, all of which are approximately 2 Debye.
 
\begin{table*}
\centering
\scriptsize
\caption{Measured column densities and upper limits for the HC$_n$O ($n$~=~3--7) family compared to the predicted values from the HC$_n$N family abundance trend (\S\ref{HCnN}) and from the chemical model (\S\ref{modeldiscussion}).  Details of the transition used for the calculation, the assumed maximum brightness temperature,  and the laboratory reference for the spectroscopy are also given.}
\begin{tabular}{c c c c c c c c}
\hline\hline
				&		\multicolumn{3}{c}{Column Density (10$^{11}$ cm$^{-2}$)}					\\
\cline{2-4}
Molecule			&		Observed			&		HC$_n$N 		&		Model		&	Transition			& $\nu$ (MHz)	&	$T_A$ (mK)		&		Ref.		\\
\hline
HC$_3$O			&		$<3$				&		200					&	0.5				&	$N_{k_a,k_c}$ = $5_{0,5} \rightarrow 4_{0,4}$, $J = 11/2 \rightarrow 9/2$, $F = 6 \rightarrow 5$	&	45327.7	&	10$^{\dagger}$	&	1,2,3	\\	
HC$_4$O			&		$<6$				&		-					&	220				&	$N_J$ = $8_{17/2} \rightarrow 7_{15/2}$, $F = 9 \rightarrow 8$							&	36494.6	&	10$^{\dagger}$	&	4	\\
HC$_5$O			&		17				&		17*					&	16				&	\multicolumn{3}{c}{See Table~\ref{transitions}}	&	5\\
HC$_6$O			&		$<10$			&		-					&	0.0003				&	$J = 27/2 \rightarrow 25/2$, $F = 14 \rightarrow 13$, $f$-parity							&	22165.2	&	10$^{\dagger}$	&	5	\\
\multirow{2}{*}{HC$_7$O}			&		\multirow{2}{*}{$\leq5$}				&		\multirow{2}{*}{10}					&	\multirow{2}{*}{5}				&	$J = 35/2 \rightarrow 33/2$									&	19204$^{\ddagger}$	&	2.4			&	5	\\
							&										&										&								&	$J = 37/2 \rightarrow 35/2$									&	20302$^{\ddagger}$	&	3.7			&	5	\\
\hline
\multicolumn{8}{l}{*Fixed to observed value.}\\
\multicolumn{8}{l}{$^{\dagger}$\citet{Kaifu:2004tk}. Accessible at www.cv.nrao.edu/SLiSE.}\\
\multicolumn{8}{l}{$^{\ddagger}$These transitions display $\Lambda$-doubling and hyperfine splitting analogous to HC$_5$O.  For simplicity, a central frequency for these splittings}\\
\multicolumn{8}{l}{is given here; individual frequencies are available in \citet{Mohamed:2005ku}.}\\
\multicolumn{8}{l}{\textbf{References.} [1]  \citet{Cooksy:1992ud} [2] \citet{Chen:1996hm} [3] \citet{Cooksy:1995bx} [4] \citet{Kohguchi:1994ci} [5] \citet{Mohamed:2005ku}}
\end{tabular}
\label{columns}
\end{table*}

\section{Discussion}
\label{discussion}

The detection of HC$_5$O, combined with the non-detections of HC$_3$O and HC$_7$O, raises the question of whether the abundance and upper limits agree with the chemistry thought to be operative in the region, and if these closely-related species might be detectable in follow-up observations.  To the former point, there are two logical avenues to explore: (1) a direct comparison to the abundance trends of the analogous carbon and nitrogen polyyne species in this source (HC$_n$N and C$_n$H) to explore if the formation and destruction pathways are perhaps similar and (2) state-of-the-art gas-grain chemical models and reaction networks.  We explore each of these below, and conclude by discussing the feasibility of future detections of other HC$_n$O radicals.

\subsection{Comparison to HC$_n$N}
\label{HCnN}

If we assume that the abundance log-linear decrease in HC$_n$O family members follows that of the HC$_n$N family reported in \citet{Loomis:2015jh}, we can predict the abundances of other HC$_n$O species.  Table~\ref{columns} shows the trend in column density of HC$_n$N mapped onto that for HC$_5$O and predicted for HC$_3$O and HC$_7$O.  Both of these values are larger than our established upper limits, particularly for HC$_3$O, where the predicted value is nearly two orders of magnitude larger.  We therefore conclude that the formation mechanisms that are responsible for HC$_n$O species are significantly different than those which govern the formation of the HC$_n$N family.

\subsection{Comparison to C$_n$H}

The presence of HC$_5$O in TMC-1 was previously predicted by \citet{Adams:1989uv}.  They propose that the dominant formation pathway for C$_n$O, HC$_n$O, and H$_m$C$_n$O molecules is through the reaction of C$_n$H$_m^+$ precursors with CO, followed by dissociative recombination.  It could therefore also be argued that the abundances may more closely follow the C$_{n-1}$H precursors.  There are, however, practical difficulties in making such a quantitative comparison, especially with respect to C$_4$H which would be the direct progenitor to HC$_5$O.  As indicated in Eq.~\ref{column}, the calculated column density is inversely proportional to $\mu^{2}$.  There is, however, significant uncertainty in the literature concerning the dipole moment of C$_4$H, as the ground state of the molecule involves a mixture of two, nearly degenerate electronic states, $^2\Sigma^+$ and $^2\Pi$, with vastly different dipole moments (0.87~--~4.3~Debye) \citep{Gratier:2016fj}.  As such, the derived abundance of C$_4$H in TMC-1 is poorly constrained, making a rigorous, quantitative analysis of this type difficult.

\begin{figure}
\centering
\includegraphics[width=0.5\textwidth]{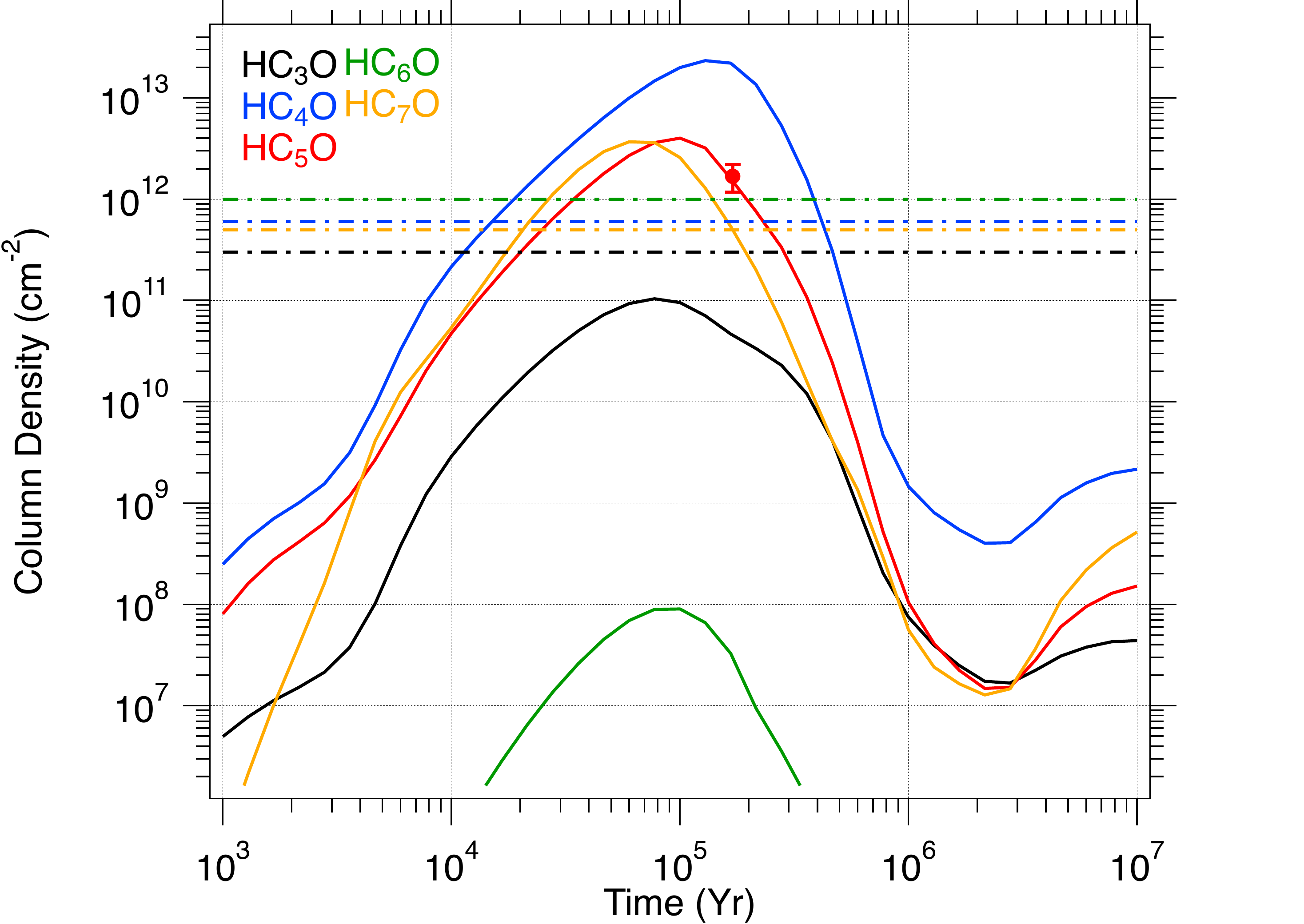}
\caption{Results of the model of the HC$_n$O radicals discussed in \S\ref{modeldiscussion}.  Gas-phase column densities predicted by the model as a function of time are shown as solid lines.  Upper limits established by observation are shown as dashed lines.  The observed column density of HC$_5$O at the best-fit cloud age ($\sim$2~$\times10^5$~yr) is indicated with a red dot.  An estimated error of 30\% is shown based on assumed flux calibration accuracy.}
\label{model}
\end{figure}

\subsection{Gas-Grain Chemical Model}
\label{modeldiscussion}

Instead, as an initial attempt to model the chemistry of the HC$_n$O series with $n=3-7$, we have used the \textsc{nautilus} 3-phase astrochemical model developed in Bordeaux \citep{Ruaud:2016bv} with the KIDA 2014 network \citep{Wakelam:2015dr}, which we have updated to include reactions related to these species. In a previous study, \citet{Adams:1989uv} predicted observable abundances of the $4\leq n \leq 6$ HC$_n$O radicals, assuming formation via the dissociative recombination reactions
\begin{equation}
  \mathrm{H}_2\mathrm{C}_n\mathrm{O}^+ + e^- \rightarrow \mathrm{HC}_n\mathrm{O} + \mathrm{H}
  \label{form1}
\end{equation}
\vspace{-1em}
\begin{equation}
  \mathrm{H}_3\mathrm{C}_n\mathrm{O}^+ + e^- \rightarrow \mathrm{HC}_n\mathrm{O} + \mathrm{H_2}.
  \label{form2}
\end{equation}

Since only H$_2$C$_3$O$^+$ and H$_3$C$_3$O$^+$ were included in the KIDA 2014 network, following \citet{Adams:1989uv}, we have added radiative association reactions \eqref{ion1} and \eqref{ion2} as formation pathways using the rate coefficients tabulated in \citet{Adams:1989uv} for the 4~$\leq$~$n$~$\leq$~$6$ HC$_n$O series and a calculated Langevin rate coefficient for reaction \eqref{ion1} leading to the HC$_7$O radical. We further updated our chemical network to include destruction of the HC$_n$O radicals by photons and ions, with the ion-polar rate coefficients calculated using the Su-Chesnavich capture approach \citep{Woon:2009ej}. 
\begin{equation}
  \mathrm{C}_{n-1}\mathrm{H}_2^+ + \mathrm{CO} \rightarrow \mathrm{H}_2\mathrm{C}_n\mathrm{O}^+ + h\nu
  \label{ion1}
\end{equation}
\vspace{-1em}
\begin{equation}
  \mathrm{C}_{n-1}\mathrm{H}_3^+ + \mathrm{CO} \rightarrow \mathrm{H}_3\mathrm{C}_n\mathrm{O}^+ + h\nu
  \label{ion2}
\end{equation}

With the updated network, simulations were run using standard TMC-1 conditions \citep{Hincelin:2011fr}, the results of which are shown in Fig.~\ref{model}. We find the best agreement with the observational results at a time of $\sim$2~$\times10^5$~yr, and previous studies have noted that TMC-1 models are in good agreement with observations at around this time \citep{Hincelin:2011fr,Majumdar:2016dn}. We note that at this cloud age, there is excellent agreement between the observed and theoretical column densities  for HC$_5$O, and values for HC$_3$O, HC$_6$O, and HC$_7$O are also below the upper limits derived in this work. At $\sim$2~$\times10^5$~yr, however, we find that HC$_4$O is overproduced by a factor of about a few compared to our present upper limit. These preliminary results suggest that the ion-neutral reactions noted in \citet{Adams:1989uv} can lead to significant abundances of HC$_n$O radicals that are in good agreement with observation at reasonable timescales. The overproduction of HC$_4$O in our preliminary models illustrates the need to further explore the chemistry of these radicals, particularly the possible importance of radical-radical reactions and of destruction pathways, which we do not consider in this work due to the current lack of theoretical studies for those pathways.

\subsection{Future Directions}

Our preliminary model well reproduces the observed column density of HC$_5$O and is in excellent agreement with the upper limits for HC$_3$O and HC$_7$O, which may indicate that the chemistry of the $n$ odd HC$_n$O radicals is rather well-constrained. If so, this would be in agreement with the composite average evidence for a population of HC$_7$O just below our detection limit (see Fig.~\ref{hc7o}b).  Given the already long integration times, and the fact that at these excitation temperatures there are no appreciably stronger transitions, however, either a significantly greater investment of observing hours or a composite average of several additional transitions would be needed to establish a firm detection.  Unless the column density for HC$_3$O is much higher than predicted by the model, it is unlikely a detection will be possible in any practical integration time.  Because our current model is unable to reproduce the observed upper limit to HC$_4$O, we reserve comment on its detectability at this juncture.

\section{Conclusions}
\label{conclusions}

We have presented the discovery of the HC$_5$O radical through observation of four well-resolved hyperfine components in high-sensitivity observations of TMC-1 with the Green Bank Telescope.  A search for other HC$_n$O radicals ($n=3-7$)   resulted in non-detections.  A first-look chemical model well-reproduces the observed column density of HC$_5$O and agrees with the upper limits inferred for HC$_3$O, HC$_6$O, and HC$_7$O.  HC$_4$O, however, is over-produced in the model, likely indicating that additional reaction pathways not considered here contribute significantly to that chemistry.  A detailed modeling study is now underway to more thoroughly examine these possibilities.  

\acknowledgments

The National Radio Astronomy Observatory is a facility of the National Science Foundation operated under cooperative agreement by Associated Universities, Inc. The Green Bank Observatory is a facility of the National Science Foundation operated under cooperative agreement by Associated Universities, Inc.  S.V.K. acknowledges support from Basic Research Program P-7 of the Presidium of the Russian Academy of Sciences.  E. H. thanks the National Science Foundation for support of his astrochemistry program.  A.M.B. is a Grote Reber Fellow, and support for this work was provided by the NSF through the Grote Reber Fellowship Program administered by Associated Universities, Inc./National Radio Astronomy Observatory.  The authors thank the anonymous referee for comments which improved the quality of this manuscript.

%\bibliography{bibliography}
%\bibliographystyle{aasjournal}

\end{document}